\documentclass[12pt]{article}%
\usepackage{amsmath}
\usepackage{graphicx}%
\usepackage{amsfonts}%
\usepackage{amssymb}
\begin{document}

\author{C. S. Unnikrishnan$^{1}$ and G. T. Gillies$^{2}$\\$^{1}$\textit{Gravitation Group, Tata Institute of Fundamental Research, }\\\textit{\ Homi Bhabha Road, Mumbai - 400 005, India}\\$^{2}$\textit{School of Engineering and Applied Science,}\\\textit{University of Virginia}, \textit{Charlottesville, VA 22904-4746, USA}\\E-mail address: $^{1}$unni@tifr.res.in, $^{2}$gtg@virginia.edu}
\title{Universality in the Gravitational Stretching of Clocks, Waves and Quantum
States }
\date{\textit{This essay received honorable mention in the Gravity Research
Foundation 2011 awards}}
\maketitle

\begin{abstract}
There are discernible and fundamental differences between clocks, waves and
physical states in classical physics. These fundamental concepts find a common
expression in the context of quantum physics in gravitational fields; matter
and light waves, quantum states and oscillator clocks become quantum
synonymous through the Planck-Einstein-de Broglie relations and the
equivalence principle. With this insight, gravitational effects on quantum
systems can be simply and accurately analyzed. Apart from providing a
transparent framework for conceptual and quantitative thinking on matter waves
and quantum states in a gravitational field, we address and resolve with
clarity the recent controversial discussions on the important issue of the
relation and the crucial difference between gravimetery using atom
interferometers and the measurement of gravitational time dilation.\bigskip


\pagebreak 

\end{abstract}

The general theory of relativity preceded the formal developments of quantum
mechanics. However, de Broglie's fundamental insight of relativistic physics
as the starting point for dealing with matter-light symmetry and for proposing
matter waves ensured that quantum dynamics of particles and light in a
gravitational field was automatically compatible in structure with general
relativity \cite{harmony,exotica}. The issue of quantum gravity -- the quantum
mechanics of gravity itself -- is of course on a different plane and it is yet
to be understood well. Gravitational quantum dynamics and physical effects are
now driving applied physics with atomic clocks and atom interferometers.

Most important of general relativistic gravitational effects, touching even
everyday life through GPS and positioning, is gravitational time dilation
\cite{einstein11}; clocks have modified rates depending on the local
gravitational potential, and the effect can be measured as the difference in
the rates of two clocks in different gravitational potentials. Gravitational
time dilation is closely linked to gravitational redshift -- the change in the
measured frequency of radiation as it moves through a gravitational potential
difference. Though the expression derived using Newtonian physics and energy
conservation, assuming the Planck relation $E=h\nu,$ and the potential
difference $\Delta\phi=gl,$ agrees with the general relativistic expression,
the correct interpretation of gravitational redshift is understood to be that
the clocks that are used to measure the frequencies at two different points
separated by distance $l$ in field $g$ run at different rates, and hence there
is difference in the frequency counts. Since frequency is a number referred to
a clock, only a change in the rate of the clock can change the frequency
\cite{einstein11}.

In classical physics a clock is not equivalent to a wave even though both are
related to periodic changes in practice. Though all waves can serve as clocks
in principle, all clocks are not waves. Also, the concept of a classical
physical state, specified by various physical quantities like momentum,
energy, angular momentum etc., has nothing to do \textit{a priori} with an
oscillator or wave. However in quantum mechanics these distinctions dissolve.
The central starting idea of quantum physics was wave-particle duality from
which emerged the concept of a wave function and so on. A stationary state of
definite energy in quantum mechanics is an `oscillator', with free time
evolution factor $\exp(-iEt/\hbar),$ with a specific frequency given by the
relation $\nu=E/h.$ Light obeys the relation $E=h\nu$ in spite of being
treated as `particles' or photons. And a clock, like an atomic clock, is based
on transitions that obey $\Delta E=h\nu.$ An important unifying idea that
emerges is that \emph{once the frequency of an oscillator is specified, its
progressive phase is equivalent to time}; there is no difference between
physical time and physical phase if an oscillator is used as the basis of time
measurement. Since the phase of an oscillator of every kind is equivalent to
time, states, photon, and matter waves can all be interpreted as `clocks' in
quantum mechanics! However, there is a price to pay for this universality --
the associated wave is an abstract and unobservable entity, manifesting only
through its relation to relevant probabilities. We stress this point because
it is important in the rigorous and correct interpretation of what measurement
of time in a gravitational field means. It is only in a space-time
interpretation of quantum physics, as in the de Brolgie-Bohm theory for
example, spatial ontological status can be ascribed to the quantum wave.

The analysis of quantum dynamics in weak (laboratory) gravitational fields
boils down to combining the preceding observation of quantum universality with
the already well known universality of the gravitational coupling. There is
just one kind of coupling of a weak gravitational field to matter:
$E_{g}=-E\phi_{g}/c^{2}$ where $E$ is total energy of the physical system. The
potential $\phi_{g}$ is simply related to the metric component $g_{00}.$ Since
there is fundamentally no distinction between general physical evolution and
quantum state evolution, general relativistic effects on a generalized quantum
oscillator, whether it is an atomic clock, an electromagnetic wave or an
unobservable quantum state oscillator, often called a matter-wave in position
representation, will follow from this interaction $E_{g}=-E\phi_{g}/c^{2}$ in
the Hamiltonian, with $E/c^{2}$ serving as an equivalent mass. We call this
universality quantum state equivalence.

Quantum state equivalence has a unifying breadth in different calculations.
For example, the general relativistic deflection of light and matter in the
gravitational field can be derived in a simple manner by appealing to
wave-particle duality and gravitational redshift \cite{deflection}. Also, the
Shapiro delay can be shown to be related to this gravitational phase delay.
The gravitational part of the quantum evolution is determined by the path
dependent integrated phase over the path given by $\Delta_{g}(x)=\int
E_{g}(x)dt/\hbar.$ For atomic clocks that work with a transition frequency
$\nu$ between two stationary states, $2\pi\nu\delta T=\Delta_{g}$ and $E=h\nu$
and the time dilation for two clocks at points $x_{1}$ and $x_{2}$ is given
by
\begin{equation}
\delta T=T\left[  \phi(x_{1})-\phi(x_{2})\right]  /c^{2}=Tgl/c^{2}%
\end{equation}
Here, $l=x_{1}-x_{2}.$ This agrees with the standard general relativistic expression.

Quantum dynamics of a system that can be in a superposition of two states with
different total energy exhibits interference between the two states with
different evolution frequencies. The process of creating the superposition of
states of different energies by giving impulses to one of the states results
in a difference in the momentum associated with the two states, and there is
state separation in position space while maintaining quantum coherence,
leading to position-energy entanglement. For example, coherent resonant
excitation by a laser with sufficient pulse duration ($\pi/2$ in terms of the
inverse Rabi frequency) that creates an equal superposition $\left(  \left|
g\right\rangle +\left|  e\right\rangle /\sqrt{2}\right)  $ of ground and
excited states $\left|  g\right\rangle $ and $\left|  e\right\rangle $
starting with the ground state, actually generates the detailed state $\left|
g,0\right\rangle +\left|  e,\hbar k\right\rangle /\sqrt{2}).$ The difference
in momentum $\hbar k$ develops into a separation of the states in space. The
quantum phase of such entangled states in a gravitational field is
particularly interesting because the coupling is universal and forms the basis
of gravimeters and inertial sensors employing atom interferometry
\cite{review-atom-inter}. For photons in a gravitational field, the relevant
coupling energy is $-h\nu\phi_{g}/c^{2}$ and for material particles it is
$-m\phi_{g}.$ Since $m\approx\left(  10^{9}-10^{11}\right)  h\nu/c^{2}$ for
neutral systems convenient for matter wave interferometry, like neutrons and
atoms, the sensitivity of matter wave interferometry is a whopping factor
$10^{10}$ higher than optical interferometry in situations involving
gravitational and inertial sensing.

For material particles, the gravitational energy is $E_{g}=-m_{g}\phi_{g}$
with the gravitational mass explicitly appearing in the quantum phase.
However, as stressed in references \cite{harmony,exotica}, there is full
compatibility with the classical equivalence principle. Since the quantum
phase is proportional to the product of this energy and the time spent in the
potential, $t\simeq l/v,$ where $\ l$ is the spatial scale and $v$ the
velocity of the particle, the accumulated phase in each path is
\begin{equation}
\Delta_{g}\simeq E_{g}t/\hbar=-m_{g}\phi_{g}l/v\hbar\label{phase1}%
\end{equation}
We can rewrite this expression, using the relation between the inertial mass
and the de Broglie wavelength in quantum theory, $\lambda_{dB}=2\pi\hbar
/m_{i}v,$ as
\begin{equation}
\Delta_{g}=-m_{g}\phi_{g}lm_{i}\lambda_{dB}/\hbar^{2} \label{phase2}%
\end{equation}
or as
\begin{equation}
\Delta_{g}=-m_{g}\phi_{g}l/m_{i}v^{2}\lambda_{dB}=-\left(  \frac{m_{g}}{m_{i}%
}\right)  \left(  \frac{E_{g}}{2E_{kin}}\right)  \left(  \frac{l}{\lambda
_{dB}}\right)  \label{phase3}%
\end{equation}
The expression is particularly interesting due to the scaling expressed in
terms of the kinetic energy and the wavelength, and more importantly due to
the appearance of the ratio of the gravitational and the inertial mass. We
note that \emph{in none of these expressions the mass term can be eliminated
without explicitly assuming the perfect validity of the Equivalence
principle}. This is crucial in the interpretation of measurements with atom
interferometer. For example, to rewrite the mass term in the equation
\ref{phase1} as a frequency one needs to assume that the gravitational mass
can be replaced with the inertial mass since all quantum frequencies refer to
the inertial mass appearing in the law of dynamics. Not recognizing this
subtle and important point can lead to misinterpretation of gravitational
effects on quantum states.

For atom interferometry involving internal states of an atom with energies
$E_{i}$, the relevant gravitational energy is $E_{g}=-(m_{g}+E_{i}/c^{2}%
)\phi_{g}.$ Usually, $E_{i}/c^{2}\ll m,$ and can be ignored. For example, in
the case of hyperfine transitions that define a typical primary clock,
$E_{i}/c^{2}\simeq10^{-15}m.$ In most situations, the phases arising from
$E_{i}\phi_{g}/c^{2}$ in the two paths are equal due to the application of a
$\pi$ pulse that inverts the quantum states in propagation resulting in the
cancellation of this contribution. Then the phase of the relevant quantum
state is $\Delta_{g}=E_{g}T=-m_{g}\phi_{g}T.$ It is important to note that
there is no reference to any `wave' or `frequency' in this expression.
However, one can write this in terms of phase shifts on \emph{fictitious waves
in real space}, with a projected and assumed spatiotemporal correspondence
with quantum states in Hilbert space in some interpretations of quantum
mechanics. The quantum state of the slow atoms is characterized by a de
Broglie wave with $\lambda=h/p\simeq h/m_{i}v\sim10^{-7}m$ in the case of cold
atom interferometers. In contrast, the Compton wavelength is $\lambda
_{c}=h/m_{i}c\sim10^{-17}m,$ which is a notional quantity and not anything
physical in the context of nonrelativistic cold atoms. It is easy to
demonstrate that the Compton frequency $\omega_{c}=m_{i}c^{2}/h$ is just a
unit conversion for mass and not relevant for the interfering wave by actually
forming the spatial fringe pattern; \emph{the fringe spacing corresponds to
the de Broglie wavelength }$\lambda=h/p$\emph{ and not to the Compton
wavelength}. The distinction becomes important and crucial if the phase shift
of the fictitious wave is interpreted as the gravitational time dilation of a
clock. The smallest time dilation factor than can be measured is $\delta
T=\Delta_{g}/\omega$ where $\omega$ is the frequency of the clock oscillator.
Therefore, the expression for the gravitational phase shift can be re-written
to obtain a seemingly exceptional sensitivity for the measurement of
gravitational time dilation if we imagine the moving atom as a real `Compton
wave clock' in space and time, in spite of the impossibility of operations
like synchronization and resetting to another standard primary clock
\cite{chu}. This is arguably much less rigorous than imagining it as a real de
Broglie wave with wavelength $h/p$ propagating in real space.

We stress the important point that a physical clock should admit standard
clock operations relative to a primary standards and for this it is necessary
that the oscillator phase is directly accessible for comparison. For the atom
interferometer gravimeter, this phase is manifested in the population of
either of the hyperfine states after recombination, determined by the
\emph{phase difference} imprinted gravitationally due to the difference in the
interaction energy, $-m(\phi_{g}(x_{1})-\phi_{g}(x_{2})).$ Hence we cannot
treat each of the individual wavepackets as individual clocks, just as the
two-state quantum superposition separated in a Stern-Gerlach magnet is not two
individual physical systems. They do not even exist in space as physical
reality, except in certain non-standard interpretations of quantum mechanics.

The geometry relevant for atom interferometry is indicated in figure 1. The
gravitational phase difference is simply the difference in phase accumulated
over the path sections A and B with spatial separation $l$ and temporal extent
$T$, with gravitational potentials $\phi_{g}(A)$ and $\phi_{g}(B).$%

\begin{figure}
[ptb]
\begin{center}
\includegraphics[
height=1.9995in,
width=3.595in
]%
{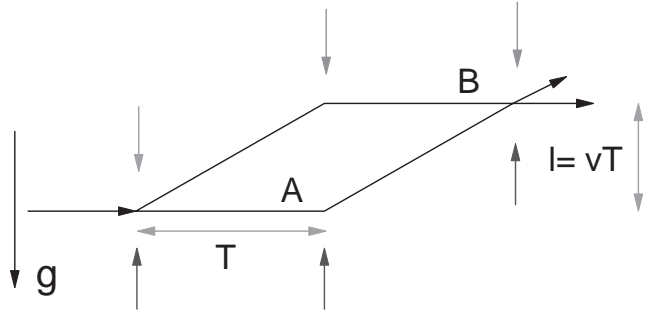}%
\caption{The space-time diagram of an atom interferometer. The laser pulses,
Pi/2, Pi and Pi/2 in NMR terminology, create the superposition of hyperfine
states that separate and recombine in space due the differential momentum
imparted.}%
\end{center}
\end{figure}

With a change of unit from mass to frequency, the phase shift is,
\begin{equation}
\Delta_{g}(x)=-m\phi_{g}(x)T/\hbar=-\frac{mc^{2}}{\hbar c^{2}}\phi
_{g}(x)T=-\omega_{c}\phi_{g}(x)T/c^{2}%
\end{equation}
Here, the coordinate $x$ distinguishes the two paths. Gravitationally rigorous
treatment \cite{ug-tobe} gives a different expression that reveals clearly the
underlying nontrivial assumption of the validity of the equivalence principle
in replacing `mass' with the Compton frequency;
\begin{equation}
\Delta_{g}(x)=-m_{g}\phi_{g}(x)T/\hbar=-\left(  \frac{m_{g}}{m_{i}}\right)
\frac{m_{i}c^{2}}{\hbar c^{2}}\phi_{g}(x)T=-\left(  \frac{m_{g}}{m_{i}%
}\right)  \omega_{c}\phi_{g}(x)T/c^{2}\label{ph-compton}%
\end{equation}
\emph{Even after assuming the equivalence principle, the appearance of \ the
`relativistic' Compton frequency is fictitious}. To see this we note that the
differential phase shift can be written as
\begin{equation}
\delta\Delta_{g}=-m\left(  \phi_{g}(x_{1})-\phi_{g}(x_{2})\right)
T/\hbar=-\frac{mv}{v}glT/\hbar=-\frac{gT^{2}}{\hbar/mv}=-2\pi gT^{2}%
/\lambda_{dB}\label{ph-deB}%
\end{equation}
$\lambda_{dB}$ is the `relative de Broglie wavelength', calculated from the
rest frame of one of the wavepackets. Since the splitting of the atomic wave
packet in the interferometer is done by pulses of light with difference in
momentum of $\hbar k_{1}-\hbar k_{2}=\hbar\kappa$ that impart a
\emph{differential recoil velocity} to the wavepackets corresponding to the
two hyperfine ground states, the relation $mv=\hbar\kappa$ holds. Hence,
$\kappa=2\pi/\lambda_{dB}.$ This determines the separation between the
wavepackets in paths A and B as $l=vT.$ In terms of the difference wave-vector
$\kappa,$
\begin{equation}
\delta\Delta_{g}=-\frac{gT^{2}}{\hbar/mv}=-\kappa gT^{2}\label{ph-kappa}%
\end{equation}
\emph{This is the expression for the phase shift in an atom
interferometer-based gravimeter}. Again, without the presumption of the
equivalence principle, this expression should be written as
\begin{equation}
\delta\Delta_{g}=-\left(  \frac{m_{g}}{m_{i}}\right)  m_{i}\left(  \phi
_{g}(x_{1})-\phi_{g}(x_{2})\right)  T/\hbar=-\left(  \frac{m_{g}}{m_{i}%
}\right)  \frac{gT^{2}}{\hbar/m_{i}v}=-\left(  \frac{m_{g}}{m_{i}}\right)
\kappa gT^{2}\label{ph-mimg}%
\end{equation}
There are several points to note. The mass term does not disappear from the
expression for the phase difference. Instead, the ratio of the gravitational
to inertial mass appears along with the relative de Broglie wavelength or
equivalently, the relative recoil momentum. The explicit dependence of phase
is on the de Broglie wavelength and not on the Compton wavelength, as revealed
in equation \ref{ph-deB}. It is clear that the gravitational phase is scaled
to the de Broglie waves, as expected for slow non-relativistic atoms, and not
to the relativistic and notional Compton wave. In fact, \emph{the phase shift
is essentially the ratio of free fall distance in the gravitational field }%
$g$\emph{ and the de Broglie wavelength,} as dictated by the equivalence
principle. In a picture projected to real space, the fringes from the
interference of the de Broglie waves will `fall' through a distance $2gT^{2}$
over time $2T$ and the phase shift is the ratio of this fringe shift and the
`centre of mass de Broglie wavelength' $h/(mv)/2=2\lambda_{dB}.$

Apart from providing a transparent analysis of the gravitational phase shift
in matter-wave gravimeters, we have clarified and resolved the recent
controversy in which M\"{u}ller \textit{et al.} claimed that gravimetry with
the atom interferometer was equivalent to a precision measurement of
gravitational redshift, the two clocks being the atomic wavepackets at A and B
\cite{chu}. If true, this amounts to a gigantic improvement in the precision
of clock comparison experiments and related tests of gravity theories.
Although Wolf \textit{et al.} noted that atoms could not be considered as
propagating Compton waves \cite{wolf}, M\"{u}ller \textit{et al.} maintained
their position invoking subtle details of possible small deviations from
general relativity \cite{chu2}. Our analysis explicitly demonstrates the
dependence on the non-relativistic momentum and the de Broglie wavelength
rather than on the Compton wavelength. The underlying assumption of the
validity of the equivalence principle in the original claim as well as in the
criticism is brought out clearly by demonstrating how the ratio of the
gravitational mass to inertial mass remains in the equations for the
differential quantum phase \cite{ug-tobe}. Moreover, we have argued that the
observable oscillator is not the Compton wave because the `fringes' are
visible only as the oscillating population in the hyperfine states or as a
spatial fringe pattern determined by the de Broglie wavelength. The Compton
wave remains as a hypothesized ghost wave with no physical manifestation.
Besides, individual wave packets do not qualify as clocks because the quantum
states they represent have no known ontological status in space, being related
to probabilities and not to actual real-time atomic positions, and they cannot
be directly synchronized and adjusted relative to another primary clock.

What constitutes a clock is an important question and hard problem in the
context of the universality of quantum states as oscillators that can respond
to gravitational potentials in proportion to their total energy. On the one
hand, all stationary quantum states have evolving phases that are modified by
local gravitational potential, and on the other, only some of them can qualify
as a genuine clock that can be compared to a primary clock standard. Our
discussion takes a physical approach with quantitative precision to provide
clarity and a satisfactory answer.

\end{document}